\pdfoutput=1

\documentclass[aps,pra,10pt,twocolumn,showpacs,superscriptaddress]{revtex4-1}
\usepackage{amsmath}
\usepackage{latexsym}
\usepackage{amssymb}
\usepackage{amsthm}
\usepackage{bm}
\usepackage{graphics,epstopdf}
\usepackage{color}\usepackage{amsmath}

\usepackage[usenames,dvipsnames,svgnames]{pstricks}
\usepackage{epsfig}
\usepackage{pst-grad} 
\usepackage{pst-plot} 
\usepackage{newlfont}
\usepackage{amssymb}
\usepackage{amsfonts}
\usepackage{amsmath}
\usepackage{amsthm}
\usepackage{gnuplottex}
\usepackage{keyval}
\usepackage{latexsym}
\usepackage{ifthen}
\usepackage{moreverb}
\usepackage{graphicx}
\usepackage{epsfig}
\usepackage{epstopdf}
\usepackage{color}
\usepackage{bm}
\usepackage[breaklinks]{hyperref}

\newtheorem{theorem}{Theorem}[section]

\usepackage{times}

\begin{document}

\title{Steerability of Quantum Coherence in Accelerated Frame}
\author{Debasis Mondal}
\email{debamondal@hri.res.in}
\author{Chiranjib Mukhopadhyay}
\email{chiranjibmukhopadhyay@hri.res.in}
\affiliation{Quantum Information and Computation Group,\\
Harish-Chandra Research Institute, Chhatnag Road, Jhunsi, Allahabad, India}


\date{\today}

\begin{abstract}
The interplay between steering and quantum coherence is studied in a scenario, where two atoms move through an external massless scalar field. We show that just like entanglement, the steering induced coherence of the equilibrium state may increase or decrease with acceleration depending on the initial condition of the state. We also investigate the condition for coherence steerability - as opposed to simple state steerability. Interestingly, we find that the quantum coherence of the equilibrium state cannot be steered, even when the steering induced coherence is non-zero. We argue that under any condition, gravity prohibits the coherence steering of the equilibrium state.
\end{abstract}

\maketitle 

\newpage

\begin{section}{Introduction}

Reconciling quantum mechanics with general relativity has been one of the mysteries yet to be completely solved by twentieth and early twenty first century theoretical physicists. Various theoretical efforts have been put forwarded to bridge this gap and several predictions have been made based on such theories. One such prediction is the so-called 'Unruh effect' \citep{unruh}\citep{davies} - which essentially tells us that an uniformly accelerated atom in an external vacuum field observes the field as a thermal bath with respect to its reference frame. The temperature of this bath is called the 'Unruh temperature' and is shown to be directly proportional to the proper acceleration of the atom. Since its original formulation for interaction with external massless scalar fields, the same effect is theoretically confirmed for much more general interactions \citep{sewell}. As the acceleration required for meaningful spontaneous excitation probability is quite high, experimentally it was not immediately confirmed. However, experimental proposals are \citep{rosu}\citep{martin} already being made. The thermalisation theorem in this context \citep{takagi} means, we can look at the atom accelerating uniformly in external field as if it is immersed in an open system and the dynamics of the atom can be studied using the master equation approach.

Study of this open system dynamics has indicated \citep{benatti} that even if two accelerating atoms immersed in an external scalar field are initially in a separable state - the final equilibrium state of the atoms obtained by tracing over the field degrees of freedom can be entangled, in general. It has also been observed \citep{benatti2} that if the atoms are assumed to be at a finite separation, then there exists a finite fixed temperature (defined in the same way as the Unruh temperature for single atoms) below which entanglement is generated immediately after the beginning of motion. This reduces to the case of vanishing separation discussed in \citep{benatti} with asymptotically non-vanishing entanglement for the equilibrium state of two atoms. It has been proved that the existence of a boundary \citep{boundary} does not change the final entanglement of the equilibrium state in this case. 

The concept of state steering, that is, manipulating the state of a part of the system by performing measurements on other entangled part of the system was initially put forwarded by Schr\"{o}dinger. There has been a major revival of the notion lately \citep{doherty}\citep{skryp}. It has been proved that state steerability is a stronger condition than the condition of the parties sharing entanglement but a weaker one than the Bell non-locality \citep{doherty}. 


Since the early days of development of quantum optics and quantum information, coherence was seen as an essentially quantum phenomenon with classicality emerging from lack of coherence \citep{zurek}. Very recently, Ref. \citep{plenio} attempted to quantify quantum coherence for arbitrary quantum states. The conditions necessary for a coherence measure are laid down in \citep{plenio}. It has been shown that the $l_{1}$ norm of coherence $C^{l_{1}}= \sum\limits_{i,j, i \neq j}  \left| \rho_{ij}\right|$, relative entropy of coherence $C_{r} = S(\rho) - S(\rho_{d})$ \citep{plenio} satisfy all these conditions and thus can be termed as coherence measures. Again, the quantum part of the uncertainty \cite{luo} or Wigner Yanase Skew Information has also been shown as an observable measure of quantum coherence \citep{girolami} or asymmetry \cite{gour1,gour2,ahmadi,marvian1,diego}. Trace distance also satisfies these conditions for a coherence measure in the qubit case and for some specified qutrit states \citep{tracenorm}. A recent research direction has been to study the interplay between quantum coherence and other resources traditionally employed in quantum information theory. The conversion between coherence and entanglement \citep{uttam}, setting up an operational resource theory of coherence \citep{winter}, role of coherence or asymmetry in setting the quantum speed limit \cite{deba} in the same way as that of entanglement \citep{wootters,manab,borras} has been achieved.
\\

 In this paper, we investigate the connection between quantum coherence and steering for two accelerating atoms in an external massless scalar field. Since steering is a stronger condition than entanglement, we expect the condition to steer the actual coherence of a system to be stricter than the condition for usual state steering. Using the criteria for the coherence steerability laid down earlier in \citep{steerability}, we have shown that in this scenario, although the equilibrium state is in general entangled as proved in \citep{benatti}, its coherence is not steerable under any initial condition. Thus, from equivalence principle, we can argue that gravity forbids such kind of non-locality.  However, coherence induced due to state steering is a different concept \citep{MID} and in this paper we show it to be non-zero and depends on the initial state of the system and the acceleration between atoms. For a family of initial conditions, we show that it is possible to have a finite acceleration, for which steering induced coherence is exactly zero - a phenomenon we term as \emph{Steering Nodes}. We also have shown that for certain different initial conditions, the steering induced coherence for large acceleration and large times (equilibrium) asymptotically approach each other. We have also investigated the effects of boundary and finite initial separation between the atoms on steering induced coherence. 
\\

The organization of our paper is as follows. In section II., we describe the dynamics of our model, first with vanishing separation between the atoms and then with the atoms placed beside an infinitely long boundary at a finite distance $z$ and a finite separation between the atoms $L$. Sections III. and IV. are dedicated to the study of steering induced coherence and coherence steerability of the equilibrium state of the atoms. We conclude in section V.
\end{section}

\begin{section}{Two Accelerated Atoms in Weak External Scalar Field}

There exists a manifestly covariant Schwinger-Tomonaga approach \citep{breuer} towards quantum evolution for relativistic  situations. But it is much complicated for our purpose. Therefore, we choose a reference frame co-moving with the system of two two-level atoms as our frame of choice.  Although we sacrifice manifest covariance, simplifications and analytical ease make up for it. In any case - if environmental correlations do not drop off significantly fast, use of a manifestly covariant formulation is limited \citep{gorini}\citep{breuer}. In the inertial frame, the two atom system as a whole follows a hyperbolic trajectory with the parameters being proper time $t$ and proper acceleration $a$. 

We assume that the atoms do not mutually interact and interact only weakly with the external scalar field. Then the Hamiltonian due to individual atoms is given by
\begin{equation}
\label{own}
H_{atom} = H_{s}^{(1)} + H_{s}^{(2)} = \frac{1}{2}\omega \sum\limits_{i =1}^{3} \vec{n}.\vec{\sigma}\otimes \mathbb{I} +  \frac{1}{2}\omega \sum\limits_{i =1}^{3} \mathbb{I} \otimes \vec{n}.\vec{\sigma}, 
\end{equation}
where $\sigma_{i}$ is the well known $i^{th}$ Pauli Matrix. Please note that natural units have been used throughout.

The contribution due to interaction of atoms with scalar field is also sum of individual atom-field interactions. If the field operators $\Phi_{\mu}$ satisfy the Klein Gordon equation, this is given by 
\begin{equation}
H_{atom-field} = \sum\limits_{\mu = 0}^{3} (\sigma_{\mu} \otimes \sigma_{0} + \sigma_{0} \otimes \sigma_{\mu}) \Phi_{\mu}(x(t)).
\end{equation}
The Field Hamiltonian will be written in terms of creation and annihilation operators, but this is not of much importance to us since we shall be integrating out the field degrees of freedom.

We shall  assume initially the two-atom system is initially in the factorized state $\rho(0) \otimes \left| 0 \rangle \langle 0\right|$ and write down the appropriate master equation  which describes the dynamics of the density matrix of the two qubit system $\rho(t)$ by tracing out the field degrees of freedom \citep{lindblad} from the well known Von Neumann equation for closed quantum system evolution. this was given \citep{kossakowski} as 
\begin{equation}
\label{master}
\frac{\partial \rho}{\partial t} = -i \left[H_{eff},\rho(t)\right] +\mathcal{L}\left[\rho(t)\right].
\end{equation}
We shall assume for simplicity that the field correlation terms are diagonal. In this case the Effective Hamiltonian has three parts :
\begin{equation}
H_{eff} = H_{1} + H_{2} + H_{12}.
\end{equation}
The first two terms correspond to \eqref{own} with the frequency $\omega$ being replaced by the renormalized frequency (say $\Omega$) and the third term is a field-generated two atom coupling term \citep{benatti}. However, at this point, let us note that this effective Hamiltonian does not involve the acceleration of the system. Since we are interested in seeing the effect of acceleration on steering induced coherence and coherence steerability - we will disregard this term and will concentrate instead on the Lindblad term, which is given by \citep{benattiprl}
\begin{eqnarray}
\label{lindblad}
\mathcal{L}\left[\rho\right] &=& \sum\limits_{i,j = 1}^{3} a_{ij}\Big{(}[ (\sigma_{j} \otimes \sigma_{0})\rho (\sigma_{i}\otimes \sigma_{0}) - \frac{1}{2} \lbrace (\sigma_{i}\sigma_{j} \otimes \sigma_{0}), \rho \rbrace] \nonumber \\
&+& [ (\sigma_{0} \otimes \sigma_{j})\rho (\sigma_{0}\otimes \sigma_{i}) - \frac{1}{2} \lbrace (\sigma_{0} \otimes \sigma_{i}\sigma_{j}) , \rho \rbrace] \nonumber \\
&+& [(\sigma_{j} \otimes \sigma_{0})\rho (\sigma_{0}\otimes \sigma_{i}) - \frac{1}{2} \lbrace (\sigma_{i} \otimes \sigma_{j}) , \rho \rbrace] \nonumber \\
&+&[ (\sigma_{0} \otimes \sigma_{j})\rho (\sigma_{i}\otimes \sigma_{0}) - \frac{1}{2} \lbrace (\sigma_{j} \otimes \sigma_{i}) , \rho \rbrace]\Big{)}, 
\end{eqnarray}
where $a_{ij}$ are elements of the Kossakowski matrix \citep{kossakowski} such that 
\begin{equation}
a_{ij} = A \delta_{ij} -iB\epsilon_{ijk}n_{k}+ C n_{i}n_{j}.
\end{equation}
Here, $A$, $B$ and $C$ are given respectively by 
\begin{equation}
\label{KossA}
A = \frac{\omega}{4 \pi} \left[\frac{1+e^{-\beta_{U} \omega}}{1- e^{-\beta_{U} \omega}} \right],
\end{equation}
\begin{equation}
\label{KossB}
B = \frac{\omega}{4\pi}
\end{equation} and
\begin{equation} 
\label{KossC}
C = \frac{\omega}{4 \pi} \left[ \frac{2}{\beta_{U}\omega} - \frac{1+e^{-\beta_{U} \omega}}{1- e^{-\beta_{U} \omega}} \right],
\end{equation}
where the Unruh temperature is given by $T_{U} = \frac{1}{\beta_{U}} = \frac{a}{2\pi}$. 
 Now if we assume the direction of $\vec{n}$ along (0,0,1) - the contribution due to the $C$ term in the Kossakowski matrix can be shown to vanish. We shall assume this in the rest of the paper unless stated otherwise.
The time dependent two atom state can now be written down in the following generic form - 
\begin{eqnarray}
\label{twoqubit}
\rho(t)  = \frac{1}{4} [ \sigma_{0} \otimes \sigma_{0} + \sum\limits_{i=1}^{3} \rho_{0i}(t) \sigma_{0}\otimes \sigma_{i} + \nonumber \\ 
\sum\limits_{i=1}^{3} \rho_{i0}(t) \sigma_{i}\otimes \sigma_{0} 
+ \sum\limits_{i=1}^{3}\sum\limits_{j=1}^{3} \rho_{ij}(t) \sigma_{i}\otimes \sigma_{j} ].
\end{eqnarray}
These time dependent coefficients can be shown to satisfy the following family of differential equations 
\begin{eqnarray}
\label{ode}
\frac{\partial \rho_{i0}(t)}{\partial t}& =& -4 A  \rho_{i0}(t) -2B(2+\tau)n_{i} +2B \sum\limits_{k=1}^{3} n_{k} \rho_{ki} \nonumber\\
\frac{\partial \rho_{0i}(t)}{\partial t} &=& -4 A  \rho_{0i}(t) -2B(2+\tau)n_{i} +2B \sum\limits_{k=1}^{3} n_{k} \rho_{ik}\nonumber\\
\frac{\partial \rho_{ij}(t)}{\partial t} &=& -4 A  \left[ 2\rho_{ij}(t)+\rho_{ji}(t) - \tau \delta_{ij} \right]+ 4B [ n_{i} \rho_{oj}(t)\nonumber \\&+& n_{j} \rho_{i0}(t)] + 2B \left[n_{i} \rho_{j0}(t) + n_{j} \rho_{0i}(t)\right] \nonumber \\
&-&2B \delta_{ij} \sum\limits_{k=1}^{3} n_{k} \left[\rho_{k0}(t) +\rho_{0k}(t) \right].
\end{eqnarray}
Finally solving these set of differential equations for equilibrium (i.e. large time $t \rightarrow \infty $), elements of the equilibrium density matrix $\rho$ can be expressed as \citep{benatti}\citep{boundary}
\begin{equation}
\rho_{0i} = \rho_{i0} = -\frac{R(\tau+3)n_{i}}{3+R^{2}}\nonumber
\end{equation}
and 
\begin{equation}
\rho_{ij} = \frac{(\tau - R^{2})\delta_{ij} + R^{2}(\tau+3)n_{i}n_{j}}{3+R^{2}},
\end{equation}
where $\tau = \Sigma_{i} \rho_{ii}(0)$ is the parameter corresponding to the choice of initial state and $R = \frac{B}{A}$ is the ratio of two constants in the Kossakowski matrix given above in \eqref{KossA} and \eqref{KossB}. From the positivity of the Kossakowski matrix - the constraint on R is $0\leqslant R \leqslant 1$ and from the positivity of the initial density matrix, the constraint on $\tau$ is $ -3 \leqslant \tau \leqslant 1$ . \\

Now we can finally explicitly write down the coefficients of \ref{twoqubit} as 

\begin{eqnarray}
\label{elements}
\rho_{01}& =& \rho_{10} = \rho_{02} = \rho_{20} = 0 \nonumber\\
\rho_{03} &= &\rho_{30} = -\frac{R(\tau +3)}{3+R^{2}}\nonumber \\
\rho_{12}&= & \rho_{21} = \rho_{13} = \rho_{31} = \rho_{23} = \rho_{32}  = 0\\
\rho_{11}&= &\rho_{22} = \frac{\tau - R^{2}}{3+R^{2}} \nonumber\\
\rho_{33}&= &\frac{\tau(1+R^{2}) + 2R^{2}}{3+R^{2}}.\nonumber
\end{eqnarray}

It was proven \citep{benatti} that this equilibrium state is in general entangled. 

\subsection*{Effect of Boundary and Finite Separation}

Let us assume a boundary at distance $z$ from atoms where the field is constrained to vanish. Let us also assume the atoms are at a finite separation $L$. This system has been solved using method of images \citep{boundary} and the analogues to Kossakowski matrix elements $A,B,C$ as earlier written in \eqref{KossA}, \eqref{KossB}, \eqref{KossC} are the following - 

\begin{eqnarray}
\label{Kossboundary}
A_{1} &=& \frac{\omega}{4\pi} \frac{1+e^{-\beta_{U}\omega}}{1-e^{-\beta_{U}\omega}} \left[1-\frac{\sin(2z\omega)}{2z\omega}\right] \nonumber \\
A_{2} &=& \frac{\omega}{4\pi} \frac{1+e^{-\beta_{U}\omega}}{1-e^{-\beta_{U}\omega}} \left[\frac{\sin(L\omega)}{L\omega}-\frac{\sin(\sqrt{L^{2}+4z^{2}}\omega)}{\sqrt{L^{2}+4z^{2}}\omega}\right] \nonumber \\ 
B_{1} &=& \frac{\omega}{4\pi} \left[1-\frac{\sin(2z\omega)}{2z\omega}\right] \nonumber \\ 
B_{2} &=& \frac{\omega}{4\pi} \left[\frac{\sin(L\omega)}{L\omega}-\frac{\sin(\sqrt{L^{2}+4z^{2}}\omega)}{\sqrt{L^{2}+4z^{2}}\omega}\right]\\ 
C_{1} &=& \frac{\omega}{4\pi} \frac{1+e^{-\beta_{U}\omega}}{1-e^{-\beta_{U}\omega}}\left[\frac{\sin(2z\omega)}{2z\omega}-1\right] \nonumber \\ 
C_{2} &=& \frac{\omega}{4\pi} \frac{1+e^{-\beta_{U}\omega}}{1-e^{-\beta_{U}\omega}} \left[\frac{-\sin(L\omega)}{L\omega}+\frac{\sin(\sqrt{L^{2}+4z^{2}}\omega)}{\sqrt{L^{2}+4z^{2}}\omega}\right]\nonumber
\end{eqnarray}

Now proceeding exactly the same way as before, the coefficients of the two qubit generic density matrix \eqref{twoqubit} for equilibrium can be written down \citep{boundary}. These are 

\begin{eqnarray}
\label{elementsboundary}
\tau &=& \frac{(2A_{1} +A_{2})B_{1}(B_{1}-B_{2})}{2A_{1}^{3}-A_{1}^{2}A_{2}-A_{2}B_{1}B{2}+A_{1}(B_{2}^{2}-A_{2}^{2})} \nonumber  \\
\rho_{0i}& =& \rho_{i0} = -\frac{(A_{1} - A_{2})B_{1}(2A_{1} + A_{2})n_{i}}{2A_{1}^{3}-A_{1}^{2}A_{2}-A_{2}B_{1}B_{2}+A_{1}(B_{2}^{2}-A_{2}^{2})}\nonumber\\
\rho_{ij} &=& \frac{(A_{1} - A_{2})B_{1}(2B_{1} + B_{2})n_{i}n_{j}}{2A_{1}^{3}-A_{1}^{2}A_{2}-A_{2}B_{1}B_{2}+A_{1}(B_{2}^{2}-A_{2}^{2})} 
\end{eqnarray} 

One interesting thing to note here is,  $\tau = \sum\limits_{i=1}^{3} \rho_{ii}$is no longer a constant of motion. Hence we can no longer use this as a parameter for the initial conditions. In the vanishing limit of $z$, we get $A_{1} = A_{2}, B_{1} = B_{2}, C_{1} = C_{2}$ and the two qubit parameters expressed in \eqref{elementsboundary} takes $\frac{0}{0}$ form. In this case, by taking proper limit, $\tau$ can be shown to be a constant of motion and the expressions for two qubit parameters reduce to those expressed in (\ref{elements}).

\end{section}



\begin{section}{Study of Steering Induced Coherence}


We consider the canonical quantum steering situation - where Alice and Bob initially share a state and Alice \emph{steers} Bob's state into a new state by performing local projective measurements on her system. In the present scenario, we consider Alice ($A$) and Bob ($B$) to be the observers on the co-moving frames of two atoms as stated above respectively.

Let Alice and Bob share the state $\rho_{AB}=\rho(t)$ \eqref{twoqubit} and let the eigenbasis of Bob's reduced state $\rho_{B} = tr_{A}(\rho_{AB})$ be the basis with respect to which coherence is calculated via different coherent measures introduced in \citep{plenio} .
\\
Clearly Bob's reduced state is initially  an incoherent state in this basis, but subsequent local projective measurements performed by Alice may change this state into a state with non-zero coherence in this basis. 
We now define the \emph{steering-induced coherence} $\bar{\mathcal{C}}$ as the maximal average coherence for Bob's new steered state which has been created by Alice performing local selective projective measurements on her part of the system. In the following, we use the $l_{1}$ norm measure of coherence to calculate this steering-induced coherence , which we shall term $\bar{\mathcal{C}}^{l_{1}}$.

At this point, we define measurement-induced disturbance (MID) which was introduced in \citep{MID} to characterize \emph{quantum} correlations -
\newtheorem*{mydef}{Definition}
\begin{mydef}
For a bipartite state $\rho_{AB}$ and local projective measurement $\Lambda^{\mathbb{E}}_{A}$($\Lambda^{\mathbb{E}}_{B}$) acting on system A(B) in $\mathbb{E}_{A}$($\mathbb{E}_{B}$) basis with the constraint that the reduced density matrix $\rho_{A}$($\rho_{B})$ remains unchanged due to that measurement - measurement induced disturbance is defined as - 
\begin{equation}
\label{twosidemid}
\mathcal{Q}(\rho_{AB}) = \inf_{\mathbb{E}_{A},\mathbb{E}_{B}} D[\rho_{AB}, \Lambda^{\mathbb{E}}_{A} \otimes \Lambda^{\mathbb{E}}_{B}(\rho_{AB})],
\end{equation}
where D[.,.] is a distance measure of our choice. 
\end{mydef}
We now define a related definition of \emph{one-sided measurement induced disturbance}.
\\
\begin{mydef} 
For a bipartite state $\rho_{AB}$ and local projective measurement $\Lambda^{\mathbb{E}}_{B}$ acting on system B in $\mathbb{E}_{B}$ basis with the constraint that the reduced density matrix $\rho_{B}$ remains unchanged due to that measurement - B-sided measurement induced disturbance is defined as - 
\begin{equation}
\label{onesidemid}
\mathcal{Q}_{B}(\rho_{AB}) = \inf_{\mathbb{E}_{B}} D[\rho_{AB}, \mathbb{I} \otimes \Lambda^{\mathbb{E}}_{B}(\rho_{AB})],
\end{equation}
where D[.,.] is a distance measure of our choice. 
\end{mydef}

It can be easily seen that this quantity is positive in general but vanishes for B-side classical states. It can be shown \emph{not} to have any coherence interpretation unlike the two sided MID introduced earlier \citep{midfan}.

Now let us formally define Steering-Induced Coherence as\citep{midfan}

\begin{mydef} 
For a bipartite quantum state $\rho$ , a local projective measurement $\xi_{i}^{A} = \left| \xi_{i}^{A} \rangle \langle \xi_{i}^{A} \right| $ by Alice  can steer Bob's state to $\rho_{B}^{\xi_{i}} = \frac{\langle \xi_{i}^{A} \left| \rho \right| \xi_{i}^{A}\rangle}{p^{\xi_{i}}}$, where the probability $p^{\xi_{i}} = Tr\left[ \rho\left(\xi_{i}^{A} \otimes \mathbb{I} \right) \right]$. Let $\mathbb{E}_{B} = \lbrace |e^{B}_{j}\rangle \rbrace$  be the eigenbasis of Bob's reduced state $\rho_{B}$. The steering-induced coherence is the maximum average coherence of Bob’s steered states on the reference basis $\mathbb{E}_{B}$ 
\begin{equation}
\label{steering}
\bar{C}(\rho) = \inf_{\mathbb{E}_{B}}\left[ \max_{\Xi_{A}} \sum\limits_{i} p_{i} C(\rho_{B}^{\xi_{i}},\mathbb{E}_{B})  \right],
\end{equation}
where the maximization is taken over all of Alice’s projective measurement basis $\Xi_{A}$ And the infimum is taken when the eigenbasis of $\rho_{B}$ i.e. $ \mathbb{E}_{B}$ is not unique.

Here $C(\rho,\Xi)$ is defined as $ C(\rho,\Xi) = D(\rho,\Lambda^{\Xi}(\rho))$ where we can take any distance measure D as we like.

\end{mydef}

Now, if the distance measure in \eqref{onesidemid} $\mathcal{Q}^{t}_{B}(\rho)$ is the well known trace norm and the steering induced coherence is written in terms of the $l_{1}$-norm as $\bar{C}^{l_{1}}$, then for a two qubit system like the one described above, the following theorem can be shown to hold \citep{midfan}  

\begin{theorem}{ }
For a two qubit state $\rho$, 
\begin{equation}
\bar{C}^{l_{1}}(\rho) = \mathcal{Q}^{t}_{B} (\rho) \nonumber
\end{equation}
\end{theorem}

\begin{figure}
{\includegraphics[scale=0.6]{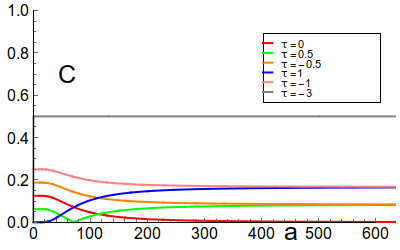}}
\caption{Steering Induced Coherence for Different Initial Conditions}
\label{Fig_Steering}
\end{figure}

Now, if we calculate the time-dependent density matrix according to the results \eqref{twoqubit}, \eqref{elements} and trace out the first system (Alice) - it can be shown that Bob's reduced density matrix is already diagonal in the computational basis. So, no further extremization is required to compute the B-sided MID vide \ref{onesidemid}. 

Graphically, the dependence of steering induced coherence on acceleration for different initial conditions based on different values of $\tau$ is depicted in Fig.\ref{Fig_Steering}.

The following properties can be observed- 

\begin{itemize}
\item For every initial condition pair $\lbrace \tau,-\tau \rbrace $ within the range of $\tau$, the  steering induced coherence for large acceleration approaches each other asymptotically and becomes a non-zero constant for each of them. 

\item For $\tau \in (0,1)$, the steering induced coherence becomes zero for some finite acceleration but then again increases with increasing acceleration. We termed these points as \emph{Steering Nodes} - the implication being, for these values of acceleration - it is not possible to induce coherence by state steering process. A  simple calculation shows that these \emph{Steering Nodes} are obtained when $\tau = R^{2}$. 

\item For $\tau \in (-3,0)$ the steering induced coherence is a monotonically decreasing function of proper acceleration $a$. However, for  $\tau \in (0,1)$ , for acceleration greater than that corresponding to the \emph{Steering Node}, the steering induced coherence is a bounded monotonically increasing function of proper acceleration $a$. 

\item The steering induced coherence of the equilibrium state increases linearly with the modulus of the initial condition parameter $\tau$ as shown in Fig. \ref{asymptotics}.
 
\end{itemize} 
\

\begin{figure}

\includegraphics[scale = 0.6]{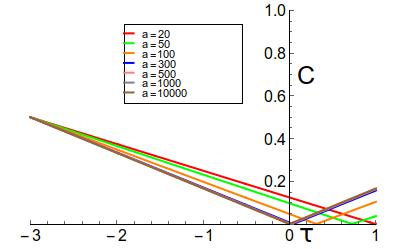}
\caption{Steering induced coherence for different values of acceleration is plotted with initial condition parameter $\tau$. We observe that $C$ is a monotonically decreasing function of $a$ for $\tau \in [-3,0)$ but it can increase for the range of $\tau \in (0,1]$ }
\label{asymptotics}
\end{figure}

\begin{subsection}*{Effect of Boundary and Finite Separation}
From \eqref{elementsboundary} , it can be easily seen that if we choose $\vec{n} = (0,0,1)$ then the equilibrium density matrix is diagonal for any finite separation or finite distance from boundary and the steering induced coherence vanishes. 

This is in line with the results in \citep{boundary} that for finite separation, entanglement between two atoms does not persist for long time.

\end{subsection}

\end{section}

\begin{section}{Investigation of Coherence Steerability}


A bipartite state is said to be steerable if and only if it does not have single system description. Such states cannot be explained by local hidden state model. Several steering inequalities have been derived using uncertainty relations, entropic uncertainty relations and fine grained uncertainty relations \cite{Jones07,Saunders, FUR_St, Walborn}. Recently, the criteria for coherence steerability for bipartite two-qubit systems were given in \citep{steerability} for different coherence measures. Intuitively, for quantum systems, it may seem that controlling the state of a system is equivalent to controlling the coherence of the system. But it was shown that for mixed states, steerability captured by different steering criteria~\cite{Saunders, FUR_St, Walborn} based on uncertainty relations are drastically different from the steerability captured by coherence property of the state \cite{steerability}. It was mentioned in \cite{steerability} that the coherence steering inequalities are the signatures of the existence of single system description of a quantum property like coherence of a state. Here, we assume that Alice and Bob are two observers in the co-moving frames of two atoms as described in the previous section and share the general two qubit state $\rho_{AB}=\rho(t)$\eqref{twoqubit}. In the following, we consider only the coherence steerability criteria with $l_{1}$ norm as the measure of quantum coherence and show that the quantum coherence of the equilibrium state \eqref{twoqubit} is unsteerable for any value of the acceleration ($a$), i.e., the state has a single system description of coherence although it has a non-zero value of steering induced coherence. 

We consider that Alice makes measurement in $\sigma_{z}$ basis and obtains binary outcome $a$ (zero or one). Bob asks Alice her measurement results and measures coherence either in $x$ or $y$ basis on his conditional state.
The Coherence of Bob's conditional state in either of these bases is given by \citep{steerability} 
\begin{equation}
C^{l_{1}}_{x(y)} \left(\rho_{B\vert \sigma_{z}^{a}} \right) =  \frac{\sqrt{\alpha_{2(1)3}^{2} + \alpha_{33}^{2}}}{1+\rho_{30}},
\end{equation}
where $\alpha_{ij} = \rho_{oi}+\rho_{ji}$.  
 
Similarly, $ C^{l_{1}}_{y(z)} \left(\rho_{B\vert \sigma_{x}^{a}} \right) =\frac{\sqrt{\alpha_{3(2)1}^{2} + \alpha_{11}^{2}}}{1+\rho_{10}} $ and  $C^{l_{1}}_{x(z)} \left(\rho_{B\vert \sigma_{y}^{a}} \right) = \frac{\sqrt{\alpha_{3(1)2}^{2} + \alpha_{22}^{2}}}{1+\rho_{20}}$ can be defined for initial measurement of Alice on $\sigma_x$ and $\sigma_y$ bases respectively.

Now the coherence steerability condition can be written as \citep{steerability} 
\begin{equation}
\label{generalsteerability}
 C^{l_{1}}_{x} \left(\rho_{B\vert \sigma_{y(z)}^{a}} \right) + C^{l_{1}}_{y} \left(\rho_{B\vert \sigma_{z(x)}^{a}} \right) +  C^{l_{1}}_{z)} \left(\rho_{B\vert \sigma_{x(y)}^{a}} \right) > \sqrt{6}.
\end{equation}

Now we use the already obtained form of the equilibrium state \eqref{elements} to get the $\alpha$-matrix as 

\begin{equation}
\label{alphamatrix}
\alpha = \begin{pmatrix}
\frac{\tau - R^{2}}{3+R^{2}}  & 0  &  0 \\
0 & \frac{\tau - R^{2}}{3+R^{2}}  &  0 \\
- \frac{R(\tau + 3)}{3+R^{2}} & - \frac{R(\tau + 3)}{3+R^{2}} &  \frac{R^{2}(\tau + 2) -R(\tau + 3) +\tau}{3+R^{2}}
\end{pmatrix}.
\end{equation}
Finally, we arrive at the condition for coherence steerability by using \eqref{generalsteerability} as
\begin{equation}
\label{steerabilitycondition}
f(\tau,R) = \frac{2(\tau - R^{2})}{3+R^{2}} + \frac{[R^{2}(\tau+2) - R(\tau+3) +\tau]}{R^{2} - R(\tau+3) +3}>\sqrt{6}.
\end{equation}

\begin{figure}
\label{coherencesteerabilityplot}
\includegraphics[scale = 0.45]{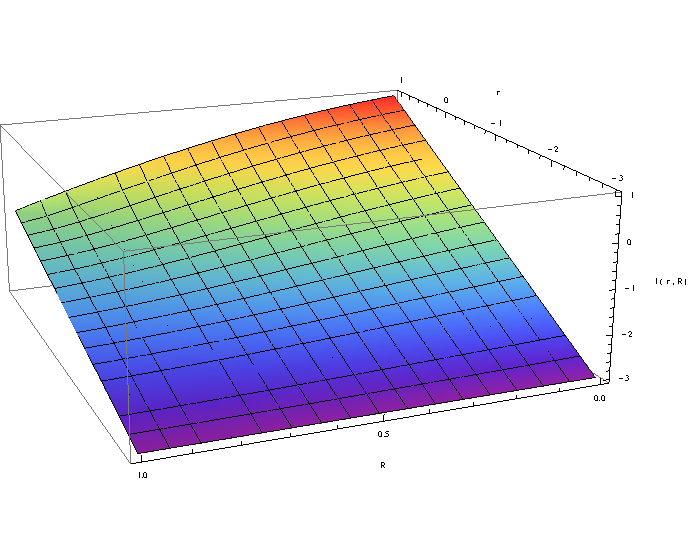}
\caption{ LHS of \eqref{steerabilitycondition} is plotted with $R$ and $\tau$ and shown never to exceed $\sqrt{6}$ - so the condition for coherence steerability is never satisfied}
\label{Fig_criteria}
\end{figure}

It can be seen from Fig \ref{Fig_criteria} that this condition is never fulfilled for allowed values of $\tau$ and $R$. Therefore, although the  equilibrium state is entangled in general, it's coherence can never be steered for any initial state. From equivalence principle, we can thus conclude that gravity also prohibits such kind of non-locality.

\begin{subsection}*{Effect of Boundary and Finite Separation}
In this case the $\alpha$-matrix as expressed earlier in \eqref{alphamatrix} is now written as 
\begin{equation}
\alpha = \begin{pmatrix}
 0  & 0  &  0 \\
0  & 0 &  0 \\
x_{1} & x_{2} & x_{3} \\
\end{pmatrix}
\end{equation}

Where $x_{1} = x_{2} = -\frac{(A_{1} - A_{2})B_{1}(2A_{1} + A_{2})}{2A_{1}^{3}-A_{1}^{2}A_{2}-A_{2}B_{1}B_{2}+A_{1}(B_{2}^{2}-A_{2}^{2})}$ and $x_{3} = \frac{(A_{1} - A_{2})B_{1}(2B_{1} + B_{2}-2A_{1} -A_{2})}{2A_{1}^{3}-A_{1}^{2}A_{2}-A_{2}B_{1}B_{2}+A_{1}(B_{2}^{2}-A_{2}^{2})}$.

Now using the condition for coherence steerability given in \citep{steerability}, we get the analogue of \eqref{steerabilitycondition} as,
\begin{equation}
\frac{x_{3}}{1+x_{1}} > \sqrt{6}
\end{equation}

A simple numerical search over large range of values of acceleration, separation $L$ and distance from  boundary $z$ shows this condition is nowhere satisfied. Therefore, the coherence is not steerable in this case also. This is true even if the initial state is fully entangled. Thus, we conclude that the frame dependent interaction between the atom and the vacuum field can create or destroy steering induced coherence like entanglement \cite{benattiprl,benatti} in the equilibrium state, although the created entanglement or correlation is not strong enough to steer the coherence of the state. 

\end{subsection}

\end{section}


\begin{section}{Conclusion}
In this paper, we have considered two accelerated atoms in a vacuum scalar field. We have shown that when the initial separation between the two atoms is zero, the steering induced coherence of the equilibrium state is non-zero and depending on the acceleration between the atoms, it can be created or destroyed like entanglement \cite{benattiprl,benatti}. But when the initial separation is finite and non-zero, the steering induced coherence decays rapidly to zero. Moreover, it is important to mention here that the boundary has no effect on steering induced coherence. Thus, from this point of view, we can conclude that the steering induced coherence is nothing but a new measure of quantum correlation. 

We also studied the coherence steerability of the equilibrium state and showed that the state cannot be steered, even when the steering induced coherence is non-zero. This phenomena can be viewed from a different angle. Suppose, two atoms are moved to two different gravitational fields from the same position. Depending on the initial state of the atoms, the steering induced coherence of the equilibrium state may increase or decrease like entanglement depending on the gravitational acceleration. But, the equilibrium state cannot be steered even when the initial state is fully entangled. Thus, we can conclude that the quantum correlation or the entanglement of the equilibrium state can never be strong enough, such that the quantum coherence of a part of the equilibrium state can be steered by the other. Gravity prohibits such non-locality. 
\end{section}

\begin{section}{Acknowledgement}
DM and CM acknowledge the research fellowship sponsored by Department of Atomic Energy, Govt. of India. 
\end{section}


\bibliographystyle{h-physrev4}

\begin{thebibliography}{60}
\bibitem{unruh} W.G. Unruh, \prd \textbf{14},870(1976).
\bibitem{davies} P.C.W. Davies, J.Phys. A. \textbf{8},609(1975).
\bibitem{sewell} G.L. Sewell, Ann. Phys. \textbf{141}, 201 (1982).
\bibitem{rosu} H.C. Rosu , Int. J. Theor. Phys. \textbf{44}, 503(2005).
\bibitem{martin}E. Martin-Martinez, I. Fuentes, R. B. Mann , \prl \textbf{107}, 131301 (2011).
\bibitem{takagi} S. Takagi, Prog. Theor. Phys. Suppl. \textbf{88},1 (1986).
\bibitem{benatti} F. Benatti , R. Floreanini, Phys. Rev. A \textbf{70},012112 (2004).
\bibitem{benatti2} F. Benatti, R. Floreanini, J. Opt. B. ; Quantum Semiclassical Opt. \textbf{7},S429(2005).
\bibitem{boundary} J. Zhang, H. Yu \pra \textbf{75}, 012101 (2007).
\bibitem{doherty} H. M. Wiseman, S. J. Jones, A. C. Doherty , \prl \textbf{98}, 140402 (2007).
\bibitem{skryp} P. Skrzypczyk, M. Navascues, D. Cavalcanti, \prl \textbf{112}, 180404 (2014).
\bibitem{zurek} W. Zurek \rmp \textbf{75},715 (2003).
\bibitem{plenio} T. Baumgratz, M. Cramer, M.B. Plenio, Phys. Rev. Lett. \textbf{113},140401 (2014).
\bibitem{luo} S. Luo, Theor. Math. Phys. {\bf 143}, 681 (2005).
\bibitem{girolami} D. Girolami \prl \textbf{113}, 170401 (2014).
\bibitem{marvian1}I. Marvian, and R. W. Spekkens, Phys. Rev. A 90, 014102 (2014).
\bibitem{gour1}G. Gour and R. W. Spekkens, New J. Phys. 10, 033023 (2008).
\bibitem{gour2}G. Gour, I. Marvian and R. W. Spekkens, Phys. Rev. A 80, 012307
 (2009).
\bibitem{ahmadi}M. Ahmadi, D. Jennings, and T. Rudolph, New J. Phys. 15, 013057
 (2013).
 \bibitem{diego}D. P. Pires, L. C. C\'eleri, D. O. Soares-Pinto, Phys. Rev. A {\bf 91}, 042330 (2015).
\bibitem{tracenorm} L. Shao, Z. Xi, H. Fan , Y. Li \pra \textbf{91}, 042120 (2015).
\bibitem{uttam} A. Streltsov, U. Singh, H.S. Dhar, M.N. Bera, G. Adesso  \prl \textbf{115}, 020403 (2015).
\bibitem{winter} A. Winter, D. Yang, arXiv 1506.07975 (2015).
\bibitem{deba}D. Mondal, C. Datta and S. Sazim, arXiv:1506.03199 (2015).
\bibitem{wootters} W. Wootters ,Phil. Trans. Roy. Soc. A \textbf{356 }, 1717 (1998).
\bibitem{manab} M. N. Bera, R. Prabhu, A K Pati, A Sen(De), U Sen, arxiv:1303.0706.
\bibitem{borras}A. Borrás, M. Casas, A. R. Plastino, and A. Plastino, Phys. Rev. A {\bf 74}, 022326 (2006).
\bibitem{steerability} D. Mondal, T. Pramanik, A.K.Pati , arXiv 1508.03770 (2015).
\bibitem{curved2} N.D. Birell, P.C.W. Davies, \emph{Quantum Fields in Curved Spaces}(Cambridge University Press, Cambridge, UK , 1982).
\bibitem{lindblad}G. Lindblad, Commun. Math. Phys. \textbf{48}, 119 (1976).
\bibitem{MID} S. Luo Phys. Rev. A  \textbf{77},022301 (2008).
\bibitem{breuer} H. P. Breuer, F. Petruccione, \emph{The Theory of Open Quantum Systems}(Oxford University Press, Oxford,2002).
\bibitem{gorini} V. Gorini, A. Frigerio, M. Verri, A. Kossakowski, E.C.G. Sudarshan , Rep. Math. Phys. \textbf{13}, 149 (1976).
\bibitem{kossakowski} V. Gorini, A. Kossakowski, E. C. G. Surdarshan, J. Math. Phys. \textbf{17}, 821 (1976).
\bibitem{benattiprl} F.Benatti, R. Floreanini, M. Piani, \prl \textbf{91}, 070402(2003).
\bibitem{midfan} X. Hu , H. Fan, arXiv 1508.01978 (2015).

\bibitem{Jones07} H. M. Wiseman, S. J. Jones, and A. C. Doherty, Phys. Rev. Lett. {\bf 98}, 140402 (2007); S. J. Jones, H. M. Wiseman, and A. C. Doherty, Phys. Rev. A {\bf 76}, 052116 (2007).
\bibitem{Saunders} D. J. Saunders, S. J. Jones, H. M. Wiseman, and G. J. Pryde, Nature Phys. {\bf 6}, 845 (2010).

\bibitem{FUR_St}  T. Pramanik, M. Kaplan, and A. S. Majumdar, Phys. Rev. A {\bf 90}, 050305(R) (2014); P. Chowdhury, T. Pramanik, A. S. Majumdar, arxiv:1503.04697.

\bibitem{Walborn} S. P. Walborn, A. Salles, R. M. Gomes, F. Toscano, and P. H. Souto Ribeiro, Phys. Rev. Lett. {\bf 106}, 130402 (2011); J. Schneeloch, C. J. Broadbent, S. P. Walborn, E. G. Cavalcanti, and J. C. Howell. Phys. Rev. A {\bf 87}, 062103 (2013).
\bibitem{maximal} X. Hu, A, Milne, B. Zhang, H. Fan , arXiv 1507.02358(2015).
\end{thebibliography}

\end{document}